\documentclass[aps,pre,twocolumn,superscriptaddress]{revtex4-1}
\usepackage[dvips]{graphicx}
\usepackage{amsmath}
\usepackage{verbatim}
\newcommand{\etalia}{{\it et al.~}}
\newcommand{\la}{\left\langle}
\newcommand{\ra}{\right\rangle}

\newcommand{\kt}{\tilde\kappa}
\newcommand{\rt}{\tilde r}

\begin{document}

\title{Structure and Osmotic Pressure of Ionic Microgel Dispersions}

\author{Mary M. Hedrick}
\affiliation{Department of Physics, North Dakota State University,
Fargo, ND 58108-6050, USA}
\affiliation{Department of Chemistry and Biochemistry, North Dakota State University,
Fargo, ND 58108-6050, USA}
\author{Jun Kyung Chung} 
\altaffiliation[Permanent address: ]{Dongmyung School Corporation, \\
793 Tongil-ro Eunpyung-gu, Seoul, Republic of Korea}
\affiliation{Department of Physics, North Dakota State University,
Fargo, ND 58108-6050, USA}
\author{Alan R. Denton}
\email[Corresponding author: ]{alan.denton@ndsu.edu}
\affiliation{Department of Physics, North Dakota State University,
Fargo, ND 58108-6050, USA}

\begin{abstract}
We investigate structural and thermodynamic properties of aqueous dispersions of 
ionic microgels -- soft colloidal gel particles that exhibit unusual phase behavior.  
Starting from a coarse-grained model of microgel macroions as charged spheres
that are permeable to microions, we perform simulations and theoretical calculations 
using two complementary implementations of Poisson-Boltzmann (PB) theory.  Within a
one-component model, based on a linear-screening approximation for effective electrostatic
pair interactions, we perform molecular dynamics simulations to compute macroion-macroion
radial distribution functions, static structure factors, and macroion contributions 
to the osmotic pressure.  For the same model, using a variational approximation 
for the free energy, we compute both macroion and microion contributions to the osmotic
pressure.  Within a spherical cell model, which neglects macroion correlations, we 
solve the nonlinear PB equation to compute microion distributions and osmotic pressures.
By comparing the one-component and cell model implementations of PB theory, we demonstrate 
that the linear-screening approximation is valid for moderately charged microgels.
By further comparing cell model predictions with simulation data for osmotic pressure,
we chart the cell model's limits in predicting osmotic pressures of salty dispersions.
\end{abstract}

\maketitle
\newpage


\section{Introduction}

Microgels are colloidal gel particles, typically 10-1000 nm in size, dispersed in,
and swollen by, a solvent.  Since their first synthesis 65 years ago~\cite{baker1949}, 
microgel dispersions have drawn interdisciplinary interest that has grown at an 
accelerating pace over the past two decades~\cite{wos}.  Experimental and theoretical 
attention has been driven as much by fundamental interest in the unusual properties 
of these soft materials as by practical interest in potential applications.  
Recent reviews~\cite{lyon-nieves-AnnuRevPhysChem2012,HydrogelBook2012,MicrogelBook2011,
pelton2000} describe prospective technologies, for example, in the chemical, biomedical, 
petroleum, and pharmaceutical industries.

Elastic, compressible, and interpenetrable gel particles can be highly sensitive to 
changes in solution conditions~\cite{flory1953,katchalsky1951,katchalsky1955,ciamarra2013,
nieves-bulk-shear-pre2011,weitz-sm2012}.  The degree of swelling can be controlled 
by adjusting temperature, pH, and chemical composition, including ionic strength.
Extreme responsiveness of microgels to their environment often leads to unique 
mechanical and rheological properties, especially at concentrations near 
close packing~\cite{nieves-jcp2003,nieves-macromol2009,dufresne2009,nieves-jcp2010,
nieves-bulk-pre2011,nieves-sm2011,weitz-jcp2012,bacchin2014}, and makes microgels
good candidates for chemical sensors and vehicles for targeted delivery and release
of drugs~\cite{hamidi2008,oh2008,oh2009,fery-AdvFunctMater2011,hoare-jcis2013}.

In recent years, well-characterized dispersions of monodisperse microgels have been 
synthesized by emulsion polymerization and cross-linking of polyelectrolytes, such as 
poly(N-isopropylacrylamide) (PNIPAM)~\cite{pelton1986,weitz-SM2008,yodh2013} or 
vinylpyridine~\cite{nieves-bulk-pre2011}.  The physical properties of polyelectrolyte 
microgel dispersions have been measured by a variety of experimental methods, 
including light scattering, small-angle neutron scattering, confocal microscopy, 
and osmometry, which have probed the connections between particle elasticity, 
osmotic pressure, structure, and phase behavior~\cite{richtering1999,wu2003,
richtering2008,lyon2007,weitz-pre2012,nieves-sm2012,schurtenberger-ZPC2012,
schurtenberger-SM2012,schurtenberger2013,nieves-pre2013,schurtenberger2014}.  

In water or other polar solvents, microgels become charged through dissociation of
counterions from the polyelectrolyte backbones.  The gel particles are then stabilized 
both sterically by dangling surface chains and electrostatically by electric charge.
Because of their permeability to solvent molecules and small ions, and the prevalence
of electrostatic forces, whose strength and range depend on degree of ionization and 
salt concentration, ionic microgels display unusual structural and thermodynamic 
properties~\cite{groehn2000,nieves-jcp2005,schurtenberger-ZPC2012} that are distinct 
from those of impermeable charged colloids~\cite{pusey1991}.

Experimental advances have motivated related theoretical and computational modeling. 
Theoretical understanding of the swelling of bulk ionic networks (gels) has a 
long history~\cite{flory1953,katchalsky1951,katchalsky1955,deGennes1979,barrat-joanny-pincus1992}.
More recent theoretical efforts have focused on effective electrostatic interactions
between microgel macroions~\cite{denton2003} and associated thermodynamic 
phase behavior~\cite{likos2011}.  For example, Gottwald~\etalia~\cite{gottwald2005}  
calculated the phase diagram of dense dispersions of microgels governed by
effective electrostatic interactions~\cite{denton2003}, using a powerful and 
elegant genetic algorithm to survey a wide variety of candidate crystal structures.  
Computer simulations of the primitive model~\cite{linse2002,holm2009} and of 
coarse-grained bead-spring models of polyelectrolytes~\cite{linse2002,holm2009,
molina2013,winkler2014} have elucidated ion distributions and structure in 
dispersions of ionic microgels.

The present work is distinct from previous modeling studies in several respects.
First, by focusing on significantly charged microgels, we highlight the influence 
of electrostatic interactions on bulk properties of ionic microgel dispersions.  
Second, we demonstrate several practical computational methods for modeling 
pair correlations and osmotic pressure, which are applicable also to other 
permeable macroions.  Third, by comparing predictions of two alternative 
implementations of the Poisson-Boltzmann theory of polyelectrolyte solutions, 
we systematically assess the validity of perturbative theoretical approximations 
and of the widely used cell model.

The remainder of the paper is organized as follows.
In Sec.~\ref{models}, we describe the physical models on which our simulations 
and calculations are based.  Starting from the primitive model of
polyelectrolytes and a uniform-sphere model of microgel macroions, we outline
derivations of the one-component and cell models of ionic microgel dispersions.
In Sec.~\ref{methods}, we describe the computational methods -- molecular
dynamics simulation and thermodynamic perturbation theory -- that we use to 
investigate physical properties of microgel dispersions.  Details of the effective
interactions required as input to the simulations and theory are deferred to 
the Appendix.  In Sec.~\ref{results}, 
we present numerical results for structural and thermodynamic properties, 
specifically the macroion-macroion radial distribution function, the
static structure factor, and the equation of state (osmotic pressure vs. microgel density).
Section~\ref{conclusions} concludes with a summary of our main results
and suggestions for future work.

\section{Models}\label{models}
\subsection{Primitive Model of Polyelectrolytes}\label{pm}

Underlying our investigations is the primitive model of 
polyelectrolytes~\cite{oosawa71,doi1988,hara93}, which idealizes the solvent as a 
uniform medium.  Applied to microgel dispersions, the model comprises spherical
macroions and microions (counterions and salt ions) dispersed in a dielectric
continuum, characterized only by a dielectric constant $\epsilon$.  
Dissociation of counterions from polyelectrolyte chains and electrostatic
screening are affected by pH and ionic strength.  Here we assume constant pH
and background ionic strength, thus fixing the macroion size and charge. 
For simplicity, we further assume that the macroions are monodisperse in
radius $a$ and valence $Z$ and that their radius is independent of macroion
concentration.  The latter assumption is consistent with experiments,
which indicate that crowding induces ionic microgel particles to de-swell 
only at concentrations approaching overlap~\cite{nieves-jcp2003}.

\begin{figure}
\begin{center}
\includegraphics[width=\columnwidth,angle=0]{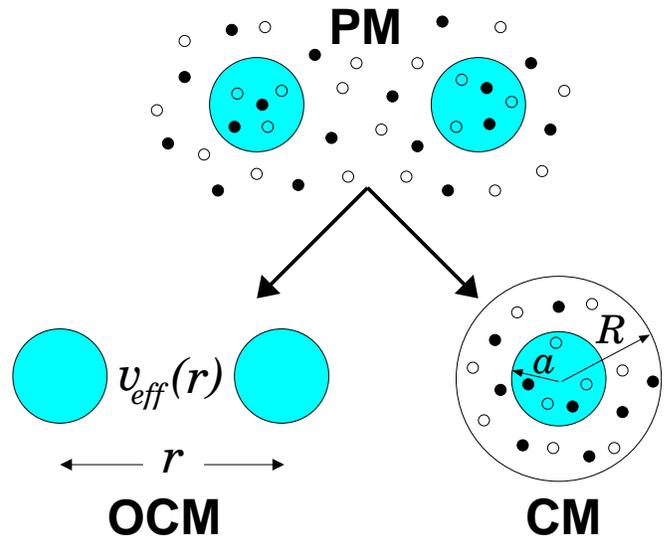}
\caption{The primitive model (PM) of ionic microgel dispersions
comprises macroions (larger spheres) and explicit counterions and coions
(smaller, light and dark spheres) dispersed in a uniform-dielectric solvent. 
The PM can be mapped onto either the coarse-grained one-component model (OCM),
with implicit microions and pseudo-macroions interacting via an effective 
pair potential $v_{\rm eff}(r)$, or the cell model (CM), consisting of a 
single macroion centered in a spherical cell along with microions.  
In all models, the solution can exchange microions (in Donnan equilibrium) 
with an implicit salt reservoir.}  
\label{fig1}
\end{center}
\end{figure}

Further abstracting, we treat the microions as monovalent point ions (valence $z=1$).
Bulk electroneutrality relates the macroion number $N_m$ to the 
counterion and coion numbers $N_{\pm}$ via $ZN_m=N_+-N_-$, assuming negative
macroions.  The total number of microions in the system can be expressed as 
$N_{\mu}=N_++N_-=ZN_m+2N_s$, where $N_s=N_-$ is the number of salt ion pairs.  
Equivalently, the corresponding ion number densities are related via $Zn_m=n_+-n_-$,
$n_{\mu}=n_++n_-$, and $n_s=n_-$.  The macroions are confined to a volume $V$, 
while the microions can exchange freely with an electrolyte reservoir of 
ion pair number density $n_0$ and bulk molar salt concentration $c_s^r=n_0/N_A$,
where $N_A$ is Avogadro's number.  Donnan equilibrium between the system and 
reservoir generally leads to a salt concentration in the system $c_s=n_s/N_A$
that is lower than in the reservoir. 

The primitive model has a long tradition in predicting bulk properties of
polyelectrolyte solutions~\cite{marcus1955} and charge-stabilized colloidal 
suspensions~\cite{pusey1991,DL1941,VO1948}, including osmotic pressure and thermodynamic 
behavior.  Nevertheless, neglect of molecular degrees of freedom, and thus 
all ion hydration effects, limits applications of the model to length scales 
much longer than the range of correlations between solvated ions.  In the systems 
that we examine here, the Debye screening length, which characterizes the typical 
ion-ion correlation length, is at least 30 nm, justifying use of the primitive model.

\subsection{Uniform-Sphere Model of Ionic Microgels}\label{usm}
Spherical microgels, consisting of cross-linked networks of polyelectrolyte chains,
typically vary in size from tens of nanometers to microns.  When dispersed in water,
the chains of ionic microgels (e.g., PNIPAM) can become charged through dissociation
of protons or other oppositely charged counterions.  The macroion size and charge
may be adjusted by varying solution conditions (e.g., temperature and pH).  The
porous structure of the gel network allows water and microions to easily penetrate.
Condensation (or close association) of some of the counterions can reduce the 
effective charge of ionic microgels, while bare Coulomb interactions between microgels
are screened by mobile microions in solution.

The simplest model of an ionic microgel represents a macroion as a uniformly charged 
sphere, penetrable to microions, characterized by only a fixed charge number density 
$n_f=3Z/(4\pi a^3)$~\cite{denton2003}.  Neglecting the internal structure of the particle,
this uniform-sphere model preserves the key properties of permeability and electrostatic 
screening, and is reasonable when the Debye screening length is much longer than the 
average distance between neighboring cross-links~\cite{barrat-joanny-pincus1992}.
It should be noted, however, that the cross-link density distribution of real microgels 
depends on the chemical composition and mode of synthesis and can be nonuniform~\cite{hoare2006}.  
From Gauss's law, the electrostatic potential energy of a microion near
a uniformly (negatively) charged spherical macroion is 
\begin{equation}
\beta v_{m\mu}(r)~=~\left\{ \begin{array} {l@{\quad\quad}l}
-\frac{\displaystyle Zz\lambda_B}{\displaystyle r}~, & r>a \\[2ex]
-\frac{\displaystyle Zz\lambda_B}{\displaystyle 2a}~
\left(3-\frac{\displaystyle r^2}{\displaystyle a^2}\right)~, &
r<a~,\end{array} \right. 
\label{vmmugel}
\end{equation}
where $\beta\equiv 1/(k_BT)$ and $\lambda_B\equiv \beta e^2/\epsilon$ is the Bjerrum length 
-- the distance at which two monovalent ions of charge $e$ interact with the typical 
thermal energy $k_BT$ at absolute temperature $T$.  This macroion-microion interaction energy,
along with the microion pair correlation functions, provides the essential input to the 
one-component model of ionic microgel dispersions with effective interactions.

\subsection{One-Component Model of Microgel Dispersions}\label{ocm}

Interactions between ionic microgel macroions comprise an electrostatic component 
and a steric component associated with the elastic nature of the gel network.  
The electrostatic interactions are modeled here by an effective-interaction theory,
previously developed by one of us~\cite{denton2003,denton-zvelindovsky2007}.
By formally integrating out the microion degrees of freedom from the partition function 
for the exact Hamiltonian of the multicomponent ion mixture, the theory maps the 
primitive model of the system onto a coarse-grained one-component model (OCM) 
governed by an {\it effective} Hamiltonian,
\begin{equation}
{\cal H}_{\rm eff}=K+E_0+U({\bf r}_1, \ldots, {\bf r}_{N_m})~.
\label{Heff}
\end{equation}
The effective Hamiltonian includes the macroion kinetic energy $K$, 
a one-body volume energy $E_0$, which accounts for the microion free energy, 
and the internal potential energy $U(\{{\bf r}_i\})$
for macroions at positions ${\bf r}_i$ ($i=1,\ldots, N_m$).
The volume energy -- specified in the Appendix -- while not affecting structural 
properties, does affect thermodynamic properties, such as pressure and phase stability.

In a linear-response approximation~\cite{denton1999,denton2000}, the microion densities 
respond linearly to the electrostatic potential of the macroions:   
\begin{equation}
\hat n_{\pm}({\bf k})~=~\chi_{\pm}(k)\hat v_{m\pm}(k)\hat n_m({\bf k}),
\qquad k\neq 0~,
\label{npmk}
\end{equation}
where $\hat n_{\pm}({\bf k})$ are Fourier transforms (with wave vector ${\bf k}$) of
the microion densities, $\chi_{\pm}(k)$ are linear response functions of a 
two-component microion plasma, and $\hat v_{m\pm}(k)$ are Fourier transforms of the 
macroion-microion electrostatic potential energy [Eq.~(\ref{vmmugel})] for positive
and negative ($z=\pm 1$) microions.  The $k\to 0$ limit must be treated separately, 
since the numbers of microions, $N_{\pm}=\hat n_{\pm}(0)$, do not respond to the 
macroion potential, but rather are determined by the electroneutrality constraint.

As described in detail elsewhere~\cite{denton1999,denton2000,denton2003}, 
the linear-response approximation [Eq.~(\ref{npmk})], which is justified for 
sufficiently weak electrostatic coupling (typically, $Z\lambda_B/a<5$), when combined
with a mean-field random-phase approximation (RPA) for the microion response
functions~\cite{denton1999,denton2000}, which is valid for weakly correlated 
monovalent ($z=1$) microions, yields the electrostatic potential 
inside and around a macroion (in units of $k_BT/e$):
\begin{equation}
\psi(r)~=~\left\{
\begin{array} {l@{\quad\quad}l}
\psi_>(r)~, & r>a \\
\psi_<(r)~, & r\leq a~,
\end{array} \right.
\label{psir}
\end{equation}
with
\begin{equation}
\psi_>(r)~=~-\frac{3Z\lambda_B}{\kt^2r}~
\left(\cosh\kt-\frac{\displaystyle \sinh\kt}
{\displaystyle \kt}\right)e^{-\kappa r}
\label{npr}
\end{equation}
and
\begin{equation}
\psi_<(r)~=~-\frac{3Z\lambda_B}{\kt^2r}~
\left[\frac{\displaystyle r}{\displaystyle a}-
\left(1+\frac{\displaystyle 1}{\displaystyle \kt}\right)
e^{-\kt}\sinh(\kappa r)\right]~, 
\label{nmr}
\end{equation}
where $\kappa=\sqrt{4\pi\lambda_B n_{\mu}}$ is the screening constant in the system
and $\kt\equiv\kappa a$.  The corresponding microion density profiles are
\begin{equation}
n_{\pm}(r)~=~n_{\infty}\mp n_{\pm}\psi(r)~,
\label{npmr}
\end{equation}
where $n_{\infty}=2n_+n_+/n_{\mu}$.

Within the linear-response approximation, many-body effective interactions vanish and
the internal potential energy can be expressed simply as a sum over macroion pairs of 
an effective pair potential $v_{\rm eff}(r_{ij})$:
\begin{equation}
U=\sum_{i<j=1}^{N_m}v_{\rm eff}(r_{ij})~,
\label{U}
\end{equation}
where $r_{ij}$ is the distance between the centers of macroions $i$ and $j$.
Within the mean-field approximation, the effective pair potential, which we take 
as input to molecular dynamics (MD) simulations and perturbation theory 
in Sec.~\ref{methods}, can be expressed as~\cite{denton2003}
\begin{equation}
v_{\rm eff}(r)~=~\left\{ \begin{array}
{l@{\quad\quad}l}
v_{\scriptscriptstyle Y}(r)~, & r > 2a \\[2ex]
v_{\rm ov}+v_{\scriptscriptstyle H}(r)~, 
& r<2a~. \end{array} \right.
\label{veff}
\end{equation}
Nonoverlapping macroions interact via an effective Yukawa (screened-Coulomb)
pair potential, 
\begin{equation}
v_{\scriptscriptstyle Y}(r)=A\frac{\displaystyle e^{-\kappa r}}{\displaystyle r}, 
\quad r>2a~, 
\label{vY}
\end{equation}
with amplitude
\begin{equation}
\beta A=\lambda_B\left[\frac{3Z}{\kt^2}\left(\cosh(\kt)
-\frac{\sinh\kt}{\kt}\right)\right]^2~.
\label{A}
\end{equation}
Overlapping macroions interact via a softer effective electrostatic pair potential 
$v_{\rm ov}(r)$ whose explicit form is given in the Appendix. 
For stronger electrostatic couplings ($Z\lambda_B/a>5$), experience with
charged colloids suggests that the Yukawa form of Eq.~(\ref{vY}) may still hold,
but with {\it renormalized} values for the parameters $Z$ and 
$\kappa$~\cite{denton2008,lu-denton2010,denton2010}.

Finally, the elastic repulsive interaction between the deformable cross-linked 
gel networks of a pair of contacting macroions we model by a Hertz 
potential~\cite{landau-lifshitz1986}
\begin{equation}
v_{\scriptscriptstyle H}(r)=B\left(1-\frac{r}{2a}\right)^{5/2}~, \quad r<2a~,
\label{vH}
\end{equation}
where the amplitude parameter $B$ governs the strength of the repulsion.
The phase diagram of Hertzian spheres was recently computed by
P\`amies \etalia~\cite{frenkel2009} using Monte Carlo simulation.
In the OCM, the total pressure, 
\begin{equation}
p=p_{\rm id}+p_0+p_m~,
\label{ptot}
\end{equation}
decomposes naturally into a macroion ideal-gas term,
\begin{equation}
p_{\rm id}=n_mk_BT~,
\label{pid}
\end{equation}
a term associated with the microion volume energy,
\begin{equation}
p_0=-\left(\frac{\partial E_0}{\partial V}\right)_{N_s/N_m}~,
\label{p01}
\end{equation}
an explicit expression for which is given in the Appendix, and a term
due to effective interactions and correlations between macroion pairs,
\begin{equation}
p_m=-\left(\frac{\partial F_m}{\partial V}\right)_{N_s/N_m}~,
\label{pm1}
\end{equation}
where $F_m$ is the macroion contribution to the excess free energy.
In Sec.~\ref{results}, we calculate the macroion pressure from simulations 
and thermodynamic theory. 

The theory of effective electrostatic interactions within the OCM, when coupled to
the RPA for the microion linear-response functions, constitutes an implementation 
of the mean-field Poisson-Boltzmann (PB) theory.  The chief appeal of this approach 
is its consistent inclusion of macroion-macroion interactions and correlations.  
Next, we review an alternative implementation of PB theory.

\subsection{Cell Model of Microgel Dispersions}\label{cm}

Another widespread implementation of PB theory is based on the cell model (CM).  
The principal advantages of this approach are its explicit incorporation of 
nonlinear microion screening and computational simplicity.  Furthermore, 
comparisons between PB theory calculations within the CM for nanogels with
uniform charge distribution and MD simulations of a more explicit bead-spring 
model demonstrate that the CM gives a reasonable representation of dilute, 
salt-free dispersions~\cite{holm2009}.
In Sec.~\ref{test}, we apply the CM to gauge the range of validity of the 
linear-screening approximation that underlies the effective pair potential 
employed in our simulations of the OCM.  First, we briefly outline the CM, 
referring the reader to a thorough review for details~\cite{deserno-holm2001}.  

Within the primitive model, the CM represents a bulk PE solution by a
single macroion, confined to a cell of like shape together with explicit 
microions and a uniform dielectric solvent.
Applied to ionic microgel dispersions, the CM places a macroion at the center
of a spherical cell, of radius $R$ determined by the macroion volume fraction
$\phi=(a/R)^3$, along with a neutralizing number of counterions and coions,
which can freely exchange with an electrolyte reservoir to maintain a fixed 
salt chemical potential (see Fig.~\ref{fig1}).
For simplicity, we assume here that the water within the macroion ($r<a$) 
has a dielectric constant equal to that of bulk water ($\epsilon\simeq 78$)
at room temperature (Bjerrum length $\lambda_B=0.7141$ nm), although 
experiments on ionic microgels indicate that the interior dielectric constant 
may be lower than in bulk~\cite{parthasarathy1996,schurtenberger-SM2012}. 

Combining the Poisson equation for the electrostatic potential $\varphi(r)$
with a Boltzmann approximation for the microion densities yields the 
nonlinear PB equation,
\begin{equation}
\psi''(r)+\frac{2}{r}\psi'(r)=\begin{cases} 
{\displaystyle \kappa_0^2\sinh\psi(r)
+\frac{3Z\lambda_B}{a^3}}~,
&0<r<a~,\\[1ex]
\kappa_0^2\sinh\psi(r)~,
&a<r<R~,
\end{cases}\label{PBeqn}
\end{equation}
where $r$ is the radial distance from the center of the cell,
$\psi(r)\equiv e\varphi(r)/k_BT$, and $\kappa_0=\sqrt{8\pi\lambda_B n_0}$ 
is the screening constant in the salt reservoir.
The mean-field Boltzmann approximation that underpins PB theory is
equivalent to the RPA invoked in most applications of response theory 
and is similarly reasonable for weakly correlated monovalent microions. 

The boundary conditions on Eq.~(\ref{PBeqn}) are as follows.  Spherical symmetry
and electroneutrality dictate that the electric field vanish at the cell center
and boundary:
\begin{equation}
\psi'_{\rm in}(0)=0~, \hspace{1cm} \psi'_{\rm out}(R)=0~,
\label{bc1}
\end{equation}
while continuity of the electrostatic potential and electric field at the 
microgel surface requires
\begin{equation}
\psi_{\rm in}(a)=\psi_{\rm out}(a)~, \hspace{1cm}
\psi'_{\rm in}(a)=\psi'_{\rm out}(a)~,
\label{bc2}
\end{equation}
the subscripts ``in" and ``out" labelling the solutions in the two regions.
By numerically solving the PB equation [Eq.~(\ref{PBeqn})], along with the
boundary conditions [Eqs.~(\ref{bc1}) and (\ref{bc2})], 
we calculate the equilibrium microion distributions within the cell.
In the process, we obtain the microion contribution to the pressure 
by applying the pressure theorem,
\begin{equation}
\beta p_{\mu}=n_+(R)+n_-(R)~,
\label{pcell}
\end{equation}
which proves to be exact in the CM~\cite{marcus1955,wennerstrom1982}.
The microion pressure $p_{\mu}$ in the CM is analogous to the 
volume pressure $p_0$ in the OCM [Eq.~(\ref{p01})].

\section{Computational Methods}\label{methods}

\subsection{Molecular Dynamics Simulations}\label{md}

\begin{figure}
\begin{center}
\includegraphics[width=\columnwidth,angle=0]{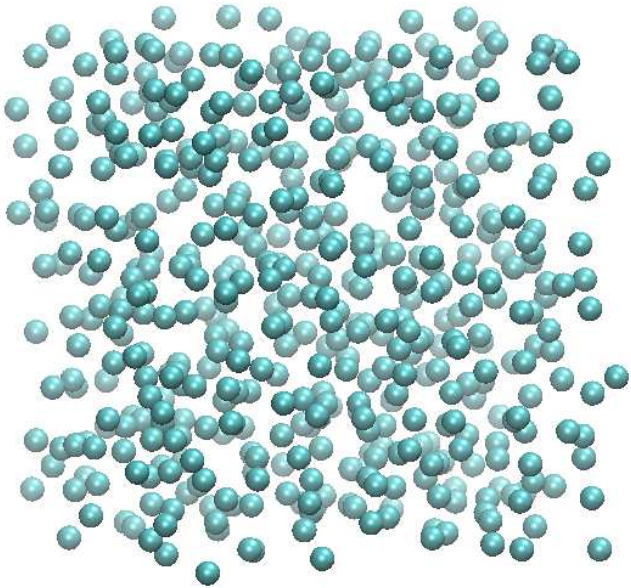}
\vspace*{-0.5cm}
\caption{Snapshot of an equilibrated fluid configuration of microgel particles
from a molecular dynamics simulation.}  
\label{snapshot}
\end{center}
\end{figure}

To explore the contribution of macroion interactions and correlations to the 
total pressure, we performed MD simulations, using LAMMPS~\cite{Plimpton1995}, 
of microgel dispersions in the OCM.  As input to the simulations, we used 
the effective pair potential described in Sec.~\ref{ocm} and the Appendix.  
Exploiting the openness of the LAMMPS source code, we coded 
the microgel pair potential [Eq.~(\ref{veff})] as a new, customized class,
supplementing the existing Yukawa pair potential class.  In the canonical ensemble
(constant $N_m$, $V$, $T$), we simulated $N_m=4000$ macroions in a cubic box with
periodic boundary conditions, initializing the particles on the sites of an fcc lattice
and cutting off the long-range Yukawa potential at a radial distance of $r_c=20/\kappa$.  
Comparisons of results from simulations of smaller systems ($N_m=500$) confirmed that
finite-size effects are negligible for $N_m=4000$.  After an initial equilibration
period of $10^6$ time steps, during which diagnostic quantities (energy, temperature, pressure)
leveled off to stable plateaus, we collected statistics for $10^7$ time steps.
Figure~\ref{snapshot} shows a typical snapshot from a simulation.

To compute the macroion contribution to the pressure from simulations of the OCM, 
we used the virial theorem.  Generalized to density-dependent 
pair potentials~\cite{louis2002,lu-denton2010}, this theorem states
\begin{equation}
p_m=\la\frac{{\cal V}_{\rm int}}{3V}\ra
-\la\left(\frac{\partial U}{\partial V}\right)_{N_s/N_m}\ra+p_{\rm tail}~,
\label{pm-sim}
\end{equation}
where ${\cal V}_{\rm int}$ denotes the internal virial, the volume derivative term
accounts for the density dependence of the effective pair potential, angular
brackets denote an ensemble average over configurations in the canonical ensemble,
and $p_{\rm tail}$ corrects for cutting off the long-range tail of the pair potential.
The internal virial is given by
\begin{equation}
{\cal V}_{\rm int}=\sum_{i=1}^{N_m}{\bf r}_i\cdot{\bf f}_i
=\sum_{i<j=1}^{N_m}r_{ij}{\rm f}_{\rm eff}(r_{ij})~,
\label{virial}
\end{equation}
where ${\bf f}_i$ is the effective force on macroion $i$ due to all other macroions
within a sphere of radius $r_c$ and ${\rm f}_{\rm eff}(r_{ij})=-v'_{\rm eff}(r_{ij})$ 
is the effective force exerted on macroion $i$ by macroion $j$.  
For nonoverlapping macroion pairs, the effective Yukawa force is
\begin{equation}
{\rm f}_{\rm eff}(r_{ij})=\left(\kappa+\frac{1}{r_{ij}}\right)
v_{\rm eff}(r_{ij})~, \quad r_{ij}>2a~.
\label{feffr>2a}
\end{equation}
An explicit expression for the effective electrostatic force between 
overlapping macroions is given in the Appendix.

The second term on the right side of Eq.~(\ref{pm-sim}) is computed,
using Eq.~(\ref{U}), as the ensemble average of
\begin{equation}
\left(\frac{\partial U}{\partial V}\right)_{N_s/N_m}
=-\frac{n_m}{V}\sum_{i<j=1}^{N_m}\left(
\frac{\partial v_{\rm eff}(r_{ij})}{\partial n_m}\right)_{N_s/N_m}~.
\label{dUdV}
\end{equation}
For nonoverlapping macroion pairs, the effective Yukawa potential [Eq.~(\ref{vY})] yields
\begin{equation}
n_m\frac{\partial v_{\rm eff}(r)}{\partial n_m}
=\left(\frac{\kt^2\sinh\kt}{\kt\cosh\kt-\sinh\kt}-3-\frac{\kappa r}{2}\right)
v_{\rm eff}(r)~.
\label{dUdVY}
\end{equation}
The corresponding expression for overlapping macroions is given in the Appendix.

Finally, the tail pressure is approximated by neglecting pair correlations 
beyond the cut-off radius, and thus setting $g(r)=1$ for $r>r_c$, with the result
\begin{eqnarray}
p_{\rm tail}&=&-\frac{2\pi}{3}n_m^2\int_{r_c}^\infty{\rm d}r\,r^3 v'_{\rm eff}(r)
\nonumber\\[1ex]
&=&\frac{2\pi}{3}n_m^2 \left(\frac{\kappa^2 r_c^2+3\kappa r_c+3}{\kappa^2}\right)
r_c v_{\rm eff}(r_c)~,
\label{ptail}
\end{eqnarray}
where the integral is evaluated using the effective Yukawa pair potential.

\subsection{Thermodynamic Perturbation Theory}\label{vfet}

To guide the choice of system parameters in our simulations, we adapt a 
thermodynamic theory previously applied with success to charged colloids.
The theory is based on a variational approximation~\cite{vanRoij1997,denton2006,hansen-mcdonald2006}
for the macroion excess free energy $F_m$, which combines first-order thermodynamic
perturbation theory with a hard-sphere (HS) reference system:
\begin{eqnarray}
&&f_m(n_m, n_s)=\min_{(d)}\left\{f_{\rm HS}(n_m, n_s;d){\phantom{\int_d^{\infty}}}
\right.
\nonumber\\[1ex]
&+&\left.2\pi n_m\int_d^{\infty}{\rm d}r\,r^2g_{\rm HS}(r, n_m;d)
v_{\rm eff}(r, n_m, n_s)\right\}.
\label{fm}
\end{eqnarray}
Here $f_m=F_m/V$ is the macroion excess free energy density, $f_{\rm HS}$ and $g_{\rm HS}$
are, respectively, the excess free energy density and pair distribution function
of the HS fluid, which we compute from the highly accurate Carnahan-Starling and 
Verlet-Weis expressions~\cite{hansen-mcdonald2006}, and $v_{\rm eff}$ is the 
effective pair potential from Sec.~\ref{md}.  Minimization of $f_m$ with respect to
the effective HS diameter $d$ yields a least upper bound to the 
free energy~\cite{hansen-mcdonald2006}.
This perturbation theory is similar to previous implementations, with
the exception that, in the present application to soft microgel macroions,
it is possible for the effective hard-sphere diameter to be smaller than the 
macroion diameter ($d<2a$).

In Donnan equilibrium, the salt concentration in the system is determined by
imposing equality of salt chemical potentials between the system and reservoir.
Treating the reservoir as an ideal gas of salt ions implies
\begin{equation}
\mu_s=2k_BT\ln(n_0\Lambda^3)~,
\label{mus1}
\end{equation}
where $\Lambda$ is the de Broglie thermal wavelength and the system salt 
chemical potential is given by
\begin{equation}
\mu_s=\left(\frac{\partial}{\partial n_s}\frac{E_0+F_m}{V}\right)_{n_m}~.
\label{mus2}
\end{equation}
An explicit expression can be found in the Appendix.
From the macroion excess free energy, an approximation for the macroion-macroion
interaction contribution to the total pressure follows immediately:
\begin{equation}
p_m=n_m^2\left(\frac{\partial f_m}{\partial n_m}\right)_{N_s/N_m}~,
\label{pm-theory}
\end{equation}
which may be compared with the corresponding expression from the
virial theorem [Eq.~(\ref{pm-sim})].  In Sec.~\ref{osmotic-pressure}, we apply 
this perturbation theory to calculate osmotic pressures of microgel dispersions.

\section{Results and Discussion}\label{results}

\subsection{One-Component Model vs.~Cell Model}\label{test}

To explore thermodynamic and structural properties of ionic microgel dispersions,
we implemented PB theory in the one-component and cell models.  
As described in Secs.~\ref{models} and \ref{methods}, we performed MD simulations 
and perturbation theory calculations in the OCM, and solved the nonlinear 
PB equation in the CM.
Our choices of system parameters were guided by the recent experimental and
modeling study of Riest \etalia~\cite{schurtenberger-ZPC2012}, who investigated
structural properties of ionic microgel dispersions characterized by $a=700$ nm, $Z=150$, 
$\kappa a=3.5$, and $B=10^4~k_BT$.  Our simulations of this model system yielded 
radial distribution functions in agreement with ref.~\cite{schurtenberger-ZPC2012},
although practically indistinguishable from those for uncharged ($Z=0$) microgels 
dispersions, suggesting that electrostatic interactions in this relatively 
weakly coupled system ($Z\lambda_B/a\simeq 0.15$) are too weak to significantly 
influence pair structure.

To amplify electrostatic effects, we considered two representative model systems 
with increased valence or decreased size of microgels: (1) $a=500$ nm, $Z=1500$, 
$c_s^r=$10~$\mu$M, and (2) $a=50$ nm, $Z=100$, $c_s^r=$100~$\mu$M.  
We refer to these systems as ``microgel" and ``nanogel" dispersions, respectively.
For both systems, we followed Riest \etalia~\cite{schurtenberger-ZPC2012} 
in fixing the amplitude of the Hertz pair potential [Eq.~(\ref{vH})] at $B=10^4~k_BT$,
corresponding to stiff elastic forces between contacting macroions. 
The chosen reservoir salt concentrations are low enough to yield relatively long 
Debye screening lengths -- for microgels, $\kappa^{-1}\simeq 100$ nm ($\kappa a\simeq 5$), and 
for nanogels, $\kappa^{-1}\simeq 30$ nm ($\kappa a\simeq 1.7$).  In such low-ionic-strength 
solutions, electrostatic interactions should significantly influence structural
and thermodynamic properties.  

\begin{figure}
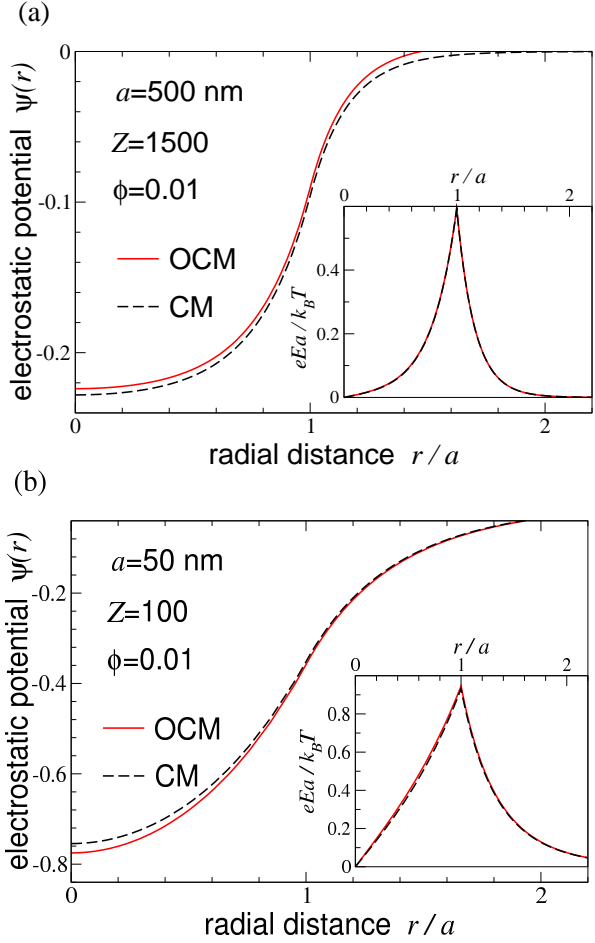

\begin{center}
\includegraphics[width=0.9\columnwidth,angle=0]{psi.z1500.a500nm.cs10.e001.eps}
\includegraphics[width=0.9\columnwidth,angle=0]{psi.z100.a50nm.cs100.e001.eps}
\vspace*{-0.2cm}
\caption{Reduced electrostatic potential $\psi(r)$ and electric field $E(r)$
(inset) predicted by linear-response theory in one-component model (OCM, solid curves) 
and nonlinear PB theory in spherical cell model (CM, dashed curves).
Results are shown for $\phi=0.01$ volume fraction dispersions of 
(a) microgels: $a=500$ nm, $Z=1500$, $c_s^r=10~\mu$M; 
and (b) nanogels: $a=50$ nm, $Z=100$, $c_s^r=100~\mu$M.
OCM $\psi(r)$ curves are offset to match CM curves at cell boundary ($r=R$).
}  
\label{ocm-vs-cm-psi1}
\end{center}
\end{figure}

According to their electrostatic coupling parameters, $Z\lambda_B/a\simeq 2.2$ 
and 1.4, both systems should still fall well within the linear-screening regime.
To confirm this assumption, however, we first compare predictions of the 
linear-response and CM implementations of PB theory.
Figures~\ref{ocm-vs-cm-psi1}-\ref{ocm-vs-cm-npm2} show numerical results for the 
electrostatic potential, electric field, and microion number density profiles
as functions of radial distance from the center of a macroion for two different 
microgel volume fractions.
The results labeled OCM are calculated from the linear-response theory
[Eq.~(\ref{npmr})], while those labeled CM are calculated by solving the 
nonlinear PB equation [Eq.~(\ref{PBeqn})] in a spherical cell.  
Although not expected to agree exactly,
because of differing boundary conditions (free vs. cell) in the two implementations,
the OCM and CM profiles are very similar, supporting the accuracy of the linearization
approximation for these parameters.  Comparison of results for $\phi=0.01$ and $0.1$ 
indicates weak sensitivity to geometrical artifacts of the CM.

Nonlinear screening may account for some part of the difference between the OCM and CM 
results, especially in the nanogel system, where a microion's electrostatic potential 
energy attains a larger fraction of the thermal energy within the macroion.  In fact,
solving the {\it linearized} PB equation does slightly alter the CM profiles,
although care is needed in choosing a consistent reference potential around which
to expand~\cite{deserno2002,denton2010}.  As an extreme test of the linear-screening
approximation, we also calculated $\psi(r)$ and $n_+(r)$ profiles for the parameters 
of ref.~\cite{holm2009}: $a\simeq 10.6$ nm, $Z=250$, $c_s^r=0$.  For this nanogel 
dispersion, we found that, compared with the CM, the OCM severely {\it underpredicts}
the counterion density inside the microgel, verifying that this strongly-coupled system 
($Z\lambda_B/a\simeq 17$) lies deep within the nonlinear regime.

\begin{figure}
\begin{center}
\includegraphics[width=0.9\columnwidth,angle=0]{npm.z1500.a500nm.cs10.e001.eps}
\includegraphics[width=0.9\columnwidth,angle=0]{npm.z100.a50nm.cs100.e001.eps}
\vspace*{-0.2cm}
\caption{Reduced microion number densities $n_{\pm}(r)$ predicted by
linear-response theory in one-component model (OCM, solid curves)
and nonlinear PB theory in spherical cell model (CM, dashed curves).
Results are shown for dispersions of (a) microgels and (b) nanogels
at volume fraction $\phi=0.01$.
System parameters are same as in Fig.~\ref{ocm-vs-cm-psi1}.
}  
\label{ocm-vs-cm-npm1}
\end{center}
\end{figure}

\begin{figure}
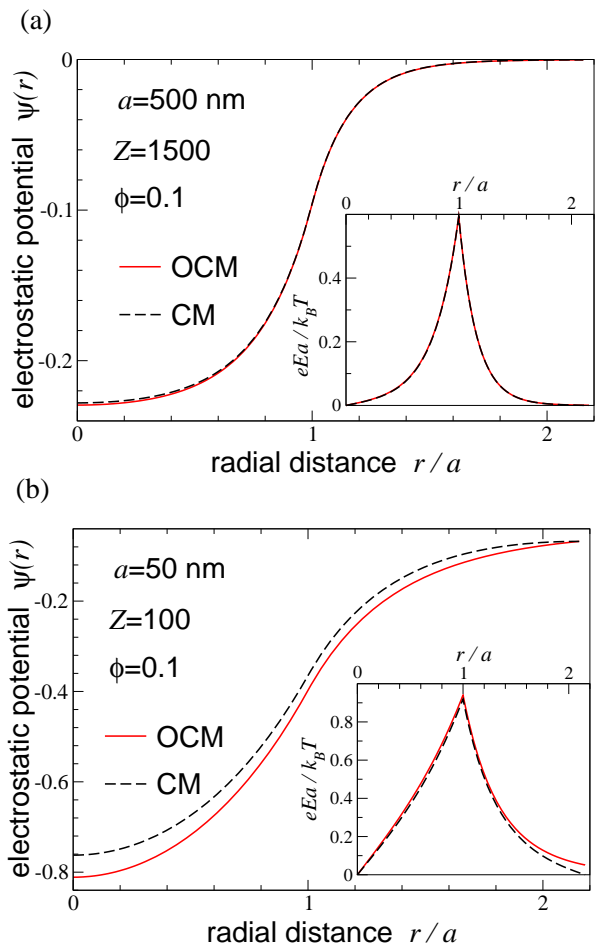

\begin{center}
\includegraphics[width=0.9\columnwidth,angle=0]{psi.z1500.a500nm.cs10.eps}
\includegraphics[width=0.9\columnwidth,angle=0]{psi.z100.a50nm.cs100.eps}
\vspace*{-0.2cm}
\caption{Reduced electrostatic potential $\psi(r)$ and electric field $E(r)$
(inset) predicted by linear-response theory in one-component model (OCM, solid curves) 
and nonlinear PB theory in spherical cell model (CM, dashed curves).
OCM $\psi(r)$ curves are offset to match CM curves at cell boundary ($r=R$).
Results are shown for dispersions of (a) microgels and (b) nanogels
at volume fraction $\phi=0.1$.
System parameters are same as in Fig.~\ref{ocm-vs-cm-psi1}.
}  
\label{ocm-vs-cm-psi2}
\end{center}
\end{figure}

\begin{figure}
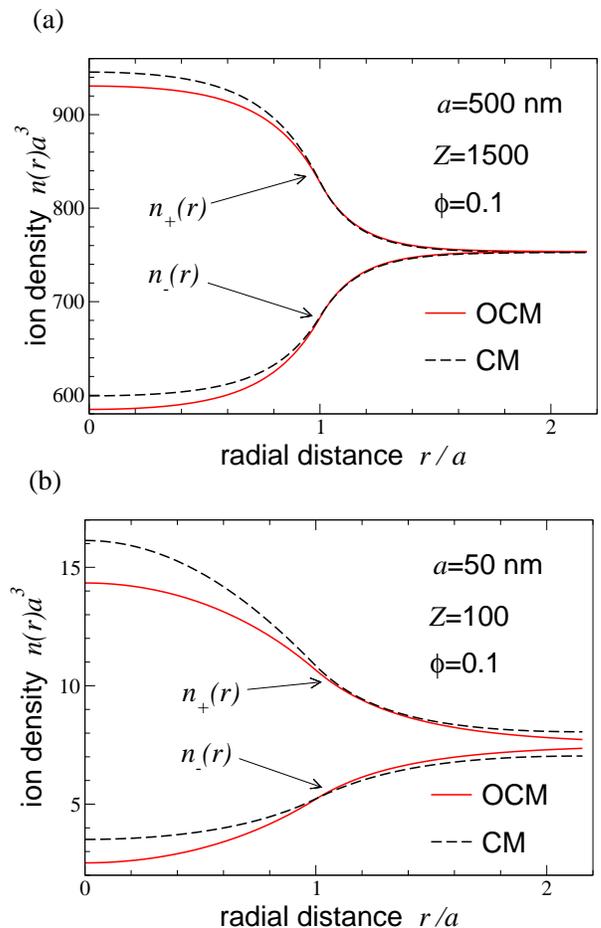

\begin{center}
\includegraphics[width=0.9\columnwidth,angle=0]{npm.z1500.a500nm.cs10.eps}
\includegraphics[width=0.9\columnwidth,angle=0]{npm.z100.a50nm.cs100.eps}
\vspace*{-0.2cm}
\caption{Reduced microion number densities $n_{\pm}(r)$ predicted by
linear-response theory in one-component model (OCM, solid curves)
and nonlinear PB theory in spherical cell model (CM, dashed curves).
Results are shown for dispersions of (a) microgels and (b) nanogels
at volume fraction $\phi=0.1$.
System parameters are same as in Fig.~\ref{ocm-vs-cm-psi1}.
}  
\label{ocm-vs-cm-npm2}
\end{center}
\end{figure}

\subsection{Osmotic Pressure of Microgel Dispersions}\label{osmotic-pressure}

Having demonstrated the accuracy of the linear-screening approximation for our 
system parameters, we proceed to input the effective pair potential
[Eq.~(\ref{veff})] into both perturbation theory and MD simulations.
Within each approach, we computed the osmotic pressure, defined as the difference
in pressure between the system and the reservoir.  To facilitate comparison,
we omit here the common macroion ideal-gas contribution [Eq.~(\ref{pid})],
defining the osmotic pressure for the two models as
\begin{equation}
\Pi~=~\left\{ \begin{array} {l@{\quad\quad}l}
p_0+p_m-p_r & {\rm (OCM)}
\\[2ex]
p_{\mu}-p_r & {\rm (CM)}~,
\end{array} \right. 
\label{Pi}
\end{equation}
where $p_r=2n_0k_BT$ is the pressure of the (ideal-gas) reservoir.
It should be noted that the volume pressure $p_0$ in the OCM, which is associated 
with the microion entropy and macroion-microion interaction energy, is 
comparable to, but distinct from, the microion pressure $p_{\mu}$ in the CM.
Within the OCM, we computed the pressure from Eqs.~(\ref{p01}) and (\ref{pm1}), 
using the linear-screening approximation for the volume energy [Eqs.~(\ref{E0})-(\ref{p02})]] 
and the variational approximation for the macroion free energy [Eq.~(\ref{fm})].
Within the CM, we computed the pressure from the pressure theorem [Eq.~(\ref{pcell})].  

\begin{figure}
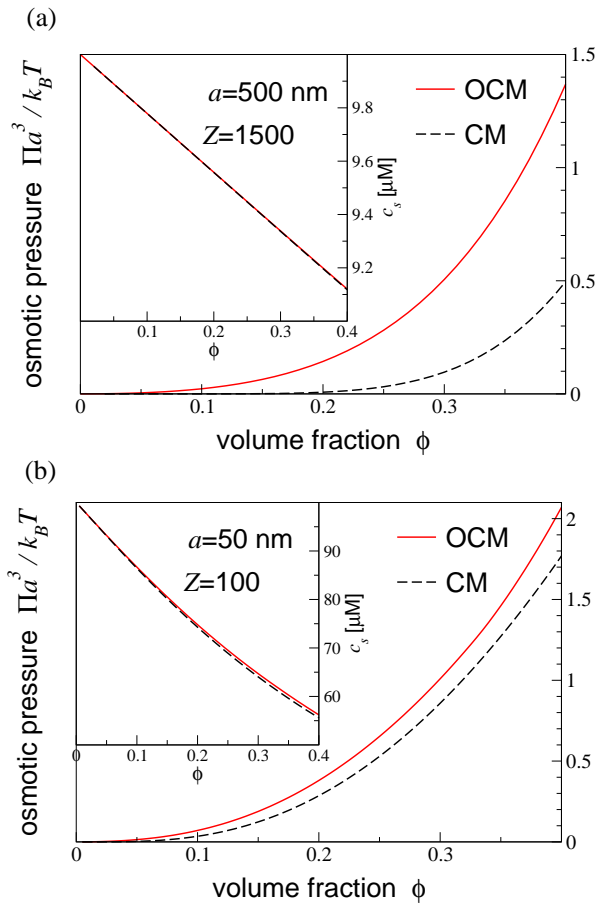

\begin{center}
\includegraphics[width=0.9\columnwidth,angle=0]{pmicrogel.z1500.a500nm.cs10.eps}
\includegraphics[width=0.9\columnwidth,angle=0]{pmicrogel.z100.a50nm.cs100.eps}
\vspace*{-0.2cm}
\caption{Osmotic pressure [Eq.~(\ref{Pi})] and salt concentration (insets) 
predicted by perturbation theory, with linearized effective interactions,
in one-component model (OCM, solid) and nonlinear PB theory in spherical cell model 
(CM, dashed).  Results are shown for dispersions of (a) microgels and (b) nanogels.
System parameters are same as in Fig.~\ref{ocm-vs-cm-psi1}.
}  
\label{pressure}
\end{center}
\end{figure}

\begin{figure}
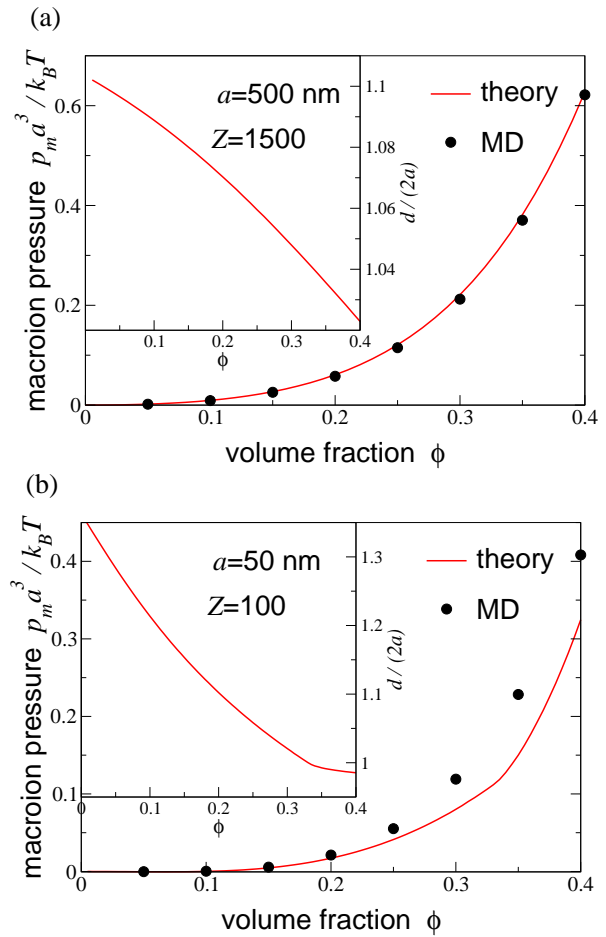

\begin{center}
\includegraphics[width=0.9\columnwidth,angle=0]{pmac.z1500.a500nm.cs10.eps}
\includegraphics[width=0.9\columnwidth,angle=0]{pmac.z100.a50nm.cs100.eps}
\vspace*{-0.2cm}
\caption{Macroion contribution to total pressure from MD simulations (symbols) 
and perturbation theory (curves), both in one-component model with linearized
effective interactions, for dispersions of (a) microgels and (b) nanogels.
System parameters are same as in Fig.~\ref{ocm-vs-cm-psi1}.
Error bars on simulation data are smaller than symbol sizes.
Insets show effective hard-sphere diameter $d$ predicted by perturbation theory.
}  
\label{pmac}
\end{center}
\end{figure}

Figure~\ref{pressure} presents a comparison between predictions of the OCM and the CM
for the osmotic pressure vs. macroion volume fraction (equation of state).
Although the system salt concentrations -- determined by equating salt chemical
potentials [Eqs.~(\ref{mus1}) and (\ref{mus2})] in Donnan equilibrium --
are nearly identical within the two models, the pressures differ significantly,
especially so for the microgel system.  Having ruled out geometry and nonlinear screening 
as significant sources of the discrepancy between predictions of the OCM and CM
for these system parameters, we seek to isolate and examine the contribution to 
the pressure originating purely from macroion correlations, which the OCM includes 
but the CM neglects.  To this end, as well as to test the accuracy of the 
variational approximation underlying the perturbation theory, we performed 
MD simulations, using the {\it same effective pair potential}, and computed the 
macroion pressure $p_m$ essentially exactly (to within statistical error).  
Comparisons between predictions of perturbation theory for $p_m$ and corresponding 
MD data are shown in Fig.~\ref{pmac}.

Good agreement between predictions of perturbation theory and results of MD
simulations demonstrates that the theory accurately models the macroion pressure.
In fact, for the microgel system, the agreement is nearly exact over the 
whole range of concentrations considered.  For the nanogel system, theory and 
simulation agree closely at lower volume fractions, while deviations emerge
and grow for $\phi>0.2$.  Given that $\kappa a$ in the nanogel system is roughly 
$1/3$ that in the microgel system, these deviations may reveal limitations of 
the perturbation theory when applied to such softly repulsive potentials.
It is also noteworthy that the deviations coincide roughly with the onset of 
significant overlap of macroions, as reflected by the drop of the effective 
hard-sphere diameter below the bare macroion diameter ($d<2a$) at $\phi\simeq 0.34$ 
(inset to Fig.~\ref{pmac}(b)).  The deviations may thus also signal the theory's 
limited ability to describe correlations between interpenetrating, soft particles.

Our analysis highlights the potentially important contribution of macroion correlations 
to the osmotic pressure of salty microgel dispersions.  These results are consistent 
with recent studies revealing limitations of the cell model implementation
of PB theory in predicting osmotic pressures of charge-stabilized colloidal 
suspensions~\cite{denton2010,hallez2014}.  While the OCM, whose thermodynamic properties 
are accessible via simulations, perturbation theory, and integral-equation 
theories~\cite{schurtenberger-ZPC2012,likos2011,hansen-mcdonald2006}, naturally includes 
the full macroion contribution to the osmotic pressure, the CM entirely neglects the 
contribution due to macroion correlations, which can be significant at higher 
salt concentrations, where the microion contribution is relatively weak.



\subsection{Structure of Microgel Dispersions}\label{structure}

\begin{figure}
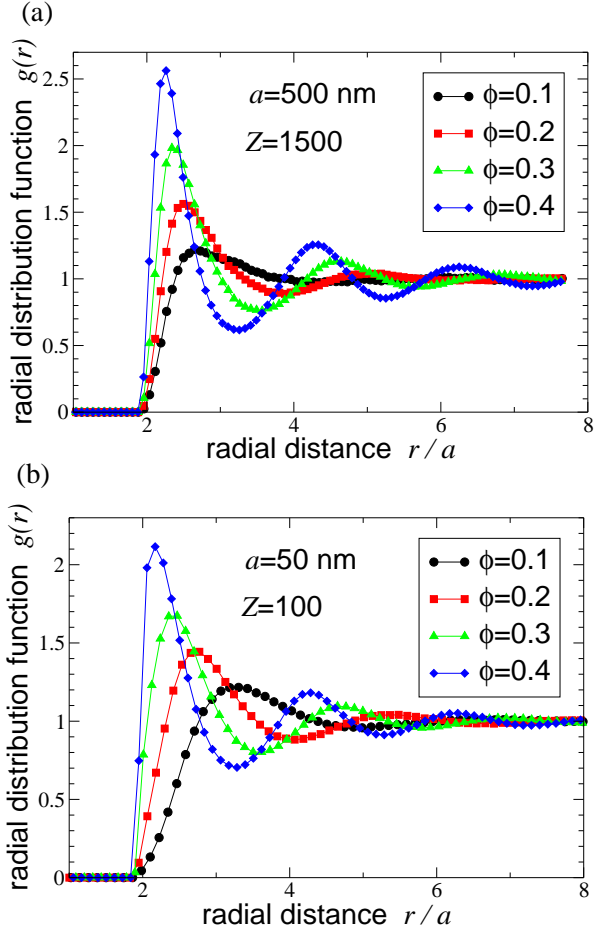

\begin{center}
\includegraphics[width=0.9\columnwidth,angle=0]{rdf.z1500.a500nm.cs10.eps}
\includegraphics[width=0.9\columnwidth,angle=0]{rdf.z100.a50nm.cs100.eps}
\vspace*{-0.2cm}
\caption{Macroion-macroion radial distribution functions $g(r)$ from 
MD simulations of the OCM for dispersions of (a) microgels and (b) nanogels.
System parameters are same as in Fig.~\ref{ocm-vs-cm-psi1}.
Radial distance $r$ is in units of the macroion radius $a$. 
Error bars are smaller than symbol sizes.  Curves are guides to the eye.
Sharpening of peaks with increasing volume fraction $\phi$ reflects strengthening 
correlations between macroions.
}  
\label{gr}
\end{center}
\end{figure}

Beyond testing and confirming the accuracy of the perturbation theory for the macroion
contribution to the osmotic pressure, simulations have the advantage of also providing
insight into structure.  To assess the significance of macroion correlations,
we calculate structural properties from MD simulations of microgel dispersions.  
As an example, Fig.~\ref{gr} shows results for the macroion-macroion radial distribution 
function $g(r)$ from our simulations of the OCM.  Data are presented for both the 
microgel and nanogel systems, and represent averages over $10^4$ configurations
(coordinate sets), spaced by intervals of $10^3$ time steps.  
We note that the hypernetted-chain (HNC) closure of the Ornstein-Zernike 
integral equation also yields very accurate results 
for these systems~\cite{schurtenberger-ZPC2012,likos2011}.

With increasing concentration, as the macroion volume fraction increases from 
$\phi=0.1$ to 0.4, the macroions evidently become more strongly correlated, as 
indicated by the growing height and narrowing width of the peaks.  This trend confirms 
the significance of macroion correlations in these concentrated fluid dispersions.
The precipitous drop of $g(r)$ near contact ($r=2a$) indicates minimal deformation of
macroions, as should be expected for such strongly repulsive (electrostatic and elastic)
pair interactions.  At the highest concentration considered here ($\phi=0.4$), however,
the nanogel macroions do interpenetrate up to a few percent of their diameter. 

\begin{figure}
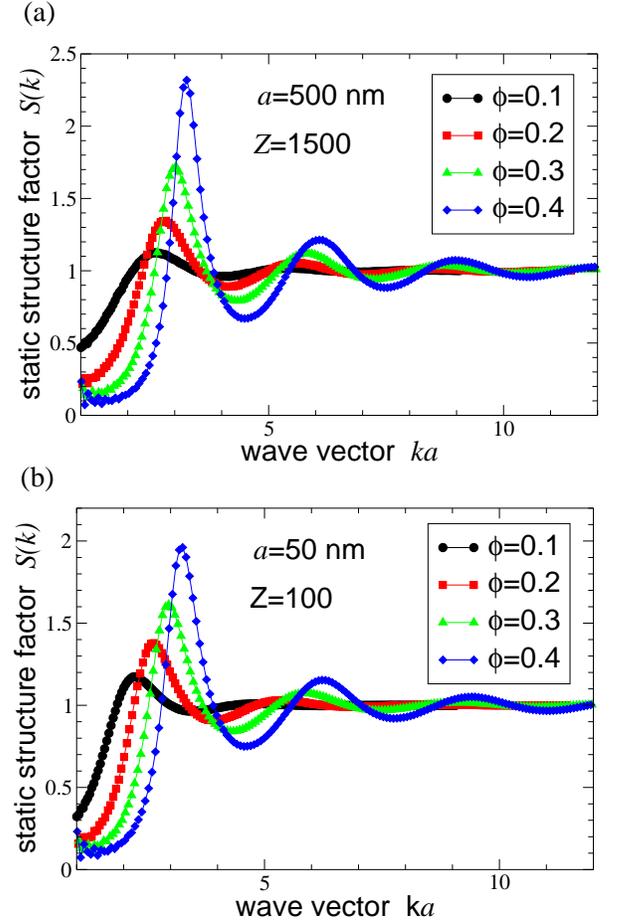

\begin{center}
\includegraphics[width=0.9\columnwidth,angle=0]{sq.z1500.a500nm.cs10.nopbc.eps}
\includegraphics[width=0.9\columnwidth,angle=0]{sq.z100.a50nm.cs100.nopbc.eps}
\vspace*{-0.2cm}
\caption{Static structure factor $S(k)$ from MD simulations of the OCM for dispersions 
of (a) microgels and (b) nanogels, corresponding to radial distribution functions 
of Fig.~\ref{gr}.  System parameters are same as in Fig.~\ref{ocm-vs-cm-psi1}.
Scattered wave vector magnitude $k$ is in units of inverse macroion radius $a^{-1}$. 
Error bars are smaller than symbol sizes and curves are guides to the eye.
Sharpening of peaks with increasing volume fraction $\phi$ reflects strengthening 
correlations between macroions.
}
\label{Sk-fig}
\end{center}
\end{figure}

To complement our results for the radial distribution function, and to
facilitate potential comparisons with scattering experiments, we also calculated 
the static structure factor, which is proportional to the Fourier transform of $g(r)$ 
and the intensity of scattered radiation (light, X-rays, neutrons).
For a uniform, isotropic fluid, the static structure factor can be defined as
\begin{equation}
S(k)=\frac{2}{N_m}\sum_{i<j=1}^{N_m}\la\frac{\sin(k r_{ij})}{k r_{ij}}\ra+1~,
\label{Sk}
\end{equation}
where $k$ is the magnitude of the scattered wave vector.  In practice, we 
calculate $S(k)$ from averages over the same configurations used to compute $g(r)$.
Although we make no attempt here to accurately calculate the long-wavelength
($k\to 0$) limit of $S(k)$, techniques to correct for finite-size effects
could be applied if needed~\cite{salacuse-1-1996,salacuse-2-1996}.

Figure~\ref{Sk-fig} shows our results for the static structure factor corresponding 
to the radial distribution functions in Fig.~\ref{gr}.  As with the $g(r)$ data,
sharpening of the peaks in $S(k)$ with increasing macroion concentration 
reflects strengthening correlations between macroions.  These correlations,
omitted from the CM, can significantly influence the bulk osmotic pressure
of salty dispersions in the fluid state~\cite{denton2010,hallez2014}, as shown
above in Sec.~\ref{osmotic-pressure}.

At sufficiently high volume fractions, or low salt concentrations, where the Debye 
screening length $\kappa^{-1}$ substantially exceeds the average nearest-neighbor separation, 
${\bar d}\sim a(\phi^{-1/3}-1)$, a solid phase may be stabilized.  In this limit,
the CM proves accurate, if only because, under these conditions, the microion contribution 
to the osmotic pressure dominates the macroion contribution.  For impenetrable charged 
colloids, the analysis of Hallez \etalia~\cite{hallez2014} shows that, for moderately 
coupled systems ($Z\lambda_B/a<8$), the CM is generally reliable for $\kappa{\bar d}<1$.
Although a similar analysis is yet to be performed for penetrable macroions, it is 
worth noting that our microgel system straddles the threshold, $\kappa{\bar d}\simeq 1$, 
while our nanogel system lies well beyond the range of accuracy of the CM ($\kappa{\bar d}>1$).

\section{Conclusions}\label{conclusions}

In summary, we have investigated thermodynamic and structural properties of ionic microgels 
-- modeled as soft, uniformly charged spheres -- dispersed in salty solutions with 
monovalent microions, using Poisson-Boltzmann theory and molecular dynamics simulation.  
We implemented PB theory within two derivatives of the primitive model:
(1) a one-component model, with effective pair interactions between microgels, and 
(2) a cell model, focused on a single microgel.  
In the OCM, we invoked a linear-screening approximation for effective electrostatic 
interactions, which we input into both MD simulations and a thermodynamic perturbation 
theory based on a variational approximation for the free energy.
In the CM, we numerically solved the nonlinear PB equation in a spherical cell geometry.  

For two model systems ranging from nanogels to microgels, with moderate electrostatic 
coupling strengths ($1<Z\lambda_B/a<3$), the linear and nonlinear implementations of 
PB theory predict very similar microion distributions, justifying the linear-screening 
approximation assumed within the OCM.  For these systems, perturbation theory and 
MD simulations -- based on the same effective pair potentials -- agree closely for 
the macroion contribution to the pressure, quantitative deviations emerging only for 
dense dispersions of interpenetrating particles with relatively long-range repulsive
interactions.  This agreement validates the variational approximation, at least for 
non-penetrating particles with short-range repulsive interactions.

Our calculations of osmotic pressure demonstrate that macroion interactions and 
correlations can make an important contribution to the total pressure of ionic 
microgel dispersions, even at relatively low (sub-mM) salt concentrations.  
The significance of macroion correlations in our model systems is confirmed by 
our MD analysis, which reveals that macroion-macroion radial distribution functions
and static structure factors are highly structured in concentrated dispersions. 
We conclude that the coarse-grained OCM provides a computationally practical
framework for exploring physical properties of ionic microgel dispersions and 
that the PB cell model, while reliable when applied to deionized solutions,
should be applied with caution to salty solutions, as was demonstrated previously
for charged colloids~\cite{denton2010,hallez2014}.

In the future, it will be important to test predictions of the OCM for thermodynamic 
and structural properties against results from experiments and from simulations of the 
primitive model, which explicitly include microions.  Previous such comparisons for 
charged colloids~\cite{denton2008,lu-denton2010,denton2010} have charted the range 
of accuracy of the OCM for impenetrable particles, identified the threshold for the 
onset of nonlinear screening effects, and helped to test and calibrate charge-renormalization 
theories.  Similar comparisons for ionic microgels would help to motivate development
of renormalization theories for dispersions of more strongly coupled (highly charged) 
penetrable particles~\cite{holm2009}.  Such a comprehensive theory would help to guide 
further exploration of the bulk modulus and phase stability of ionic microgel dispersions.
Future work may also extend the methods described here to microgels with nonuniform 
fixed charge distributions, such as core-shell particles, and to other soft ionic colloids,
such as polyelectrolyte stars, dendrites, and microcapsules~\cite{tang-denton2014}.

\acknowledgments
We thank Qiyun Tang and Brandon J.~Johnson for helpful discussions.  This work 
was supported by the National Science Foundation (Grant No.~DMR-1106331).
MMH thanks the McNair Scholars Program for support.

\appendix
\section{Effective Interaction Theory}
In the linear-screening approximation~\cite{denton2003} for the uniform-sphere model 
of microgels, the volume energy per macroion is given by
\begin{eqnarray}
\frac{\beta E_0}{N_m}&=&\frac{\beta F_p}{N_m}-3Z^2\frac{\lambda_B}{a}
\left\{\frac{1}{5}-\frac{1}{2\kt^2}+\frac{3}{4\kt^3}
\left[1-\frac{1}{\kt^2}\right.\right.
\nonumber\\
&+&\left.\left.\left(1+\frac{2}{\kt}+\frac{1}{\kt^2}\right)
e^{-2\kt}\right]\right\}
-\frac{Z}{2}\frac{n_+-n_-}{n_++n_-}~,
\label{E0}
\end{eqnarray}
where $\kt\equiv\kappa a$ and the microion plasma free energy,
\begin{equation}
\beta F_p~=~N_+[\ln(n_+\Lambda^3)-1]~+~N_-[\ln(n_-\Lambda^3)-1]~,
\label{Fplasma}
\end{equation}
is the free energy of an ideal gas of microions in a uniform compensating background.
The corresponding volume pressure [Eq.~(\ref{p01})] is given by
\begin{eqnarray}
\beta p_0&=&n_{\mu}+\frac{3}{2}Z^2\frac{\lambda_B}{a}n_m\left[-\frac{1}{\kt^2}
+\frac{9}{4\kt^3}-\frac{15}{4\kt^5}
\right.
\nonumber\\[1ex]
&+&\left.\left(\frac{3}{2\kt^2}+\frac{21}{4\kt^3}+\frac{15}{2\kt^4}+\frac{15}{4\kt^5}\right)
e^{-2\kt}\right]~,
\label{p02}
\end{eqnarray}
and the system salt chemical potential [Eq.~(\ref{mus2})] by
\begin{eqnarray}
\beta\mu_s&=&\beta\left(\frac{\partial f_m}{\partial n_s}\right)_{n_m}
+\ln(n_+\Lambda^3)+\ln(n_-\Lambda^3)
\nonumber\\[1ex]
&+&\frac{Zn_m}{n_{\mu}^2}+3Z^2\frac{\lambda_B}{a}\frac{n_m}{n_{\mu}}
\left[-\frac{1}{\kt^2}+\frac{9}{4\kt^3}-\frac{15}{4\kt^5} 
\right.
\nonumber\\[1ex]
&+&\left.\left(\frac{3}{2\kt^2}+\frac{21}{4\kt^3}
+\frac{15}{2\kt^4}+\frac{15}{4\kt^5}\right)e^{-2\kt}\right]~.
\label{mus3}
\end{eqnarray}

The effective electrostatic pair potential between overlapping macroions is 
\begin{equation}
v_{\rm ov}(r)=v_{mm}(r)+v_{\rm ind}(r)~, \quad r<2a~,
\label{vr<2a-gel}
\end{equation}
where
\begin{equation}
\beta v_{mm}(r)=Z^2\frac{\lambda_B}{a}\left(\frac{6}{5}-\frac{1}{2}\rt^2
+\frac{3}{16}\rt^3-\frac{1}{160}\rt^5\right)
\label{vmmr<2a-gel}
\end{equation}
is the bare (Coulomb) pair potential between two overlapping, uniformly charged
spheres and
\begin{eqnarray}
&&\beta v_{\rm ind}(r)=
-\left(\frac{3Z}{\kt^2}\right)^2\frac{\lambda_B}{2r}\left[
\left(1+\frac{1}{\kt}\right)^2e^{-2\kt}\sinh(\kappa r)\right.
\nonumber\\[1ex]
&&+\left(1-\frac{1}{\kt^2}\right)
\left(1-e^{-\kappa r}+\frac{1}{2}\kappa^2r^2+\frac{1}{24}\kappa^4r^4\right)
\nonumber\\[1ex]
&&-\left.\frac{2}{3}\kt^2\left(1-\frac{2}{5}\kt^2\right)\rt
-\frac{1}{9}\kt^4\rt^3-\frac{1}{720}\kt^4\rt^6 \right]
\label{vindr<2a-gel}
\end{eqnarray}
is the microion-induced pair potential, where $\rt\equiv r/a$.
The corresponding effective electrostatic force between a pair of 
overlapping macroions is given by
\begin{equation}
{\rm f}_{\rm ov}(r)=-v_{\rm ov}'(r)={\rm f}_{mm}(r)+{\rm f}_{\rm ind}(r)~,
\label{fr<2a-gel}
\end{equation}
where 
\begin{equation}
\beta {\rm f}_{mm}(r)=Z^2\frac{\lambda_B}{a^2}\left(\rt-\frac{9}{16}\rt^2+\frac{1}{32}\rt^4\right)
\label{fmmr<2a-gel}
\end{equation}
is the bare electrostatic force and 
\begin{eqnarray}
&&\beta {\rm f}_{\rm ind}(r)=
\left(\frac{3Z}\kt^2\right)^2\frac{\kappa\lambda_B}{2r}\left[
\left(1+\frac{1}\kt\right)^2e^{-2\kt}\cosh(\kappa r)
\right.
\nonumber\\[1ex]
&&+\left(1-\frac{1}{\kt^2}\right)\left(e^{-\kappa r}+\kappa r
+\frac{1}{6}\kappa^3 r^3\right)-\frac{2}{3}\kt\left(1-\frac{2}{5}\kt^2\right)
\nonumber\\[1ex]
&&-\left.\frac{1}{3}\kt^3\rt^2-\frac{1}{120}\kt^3\rt^5\right]
+\frac{\beta v_{\rm ind}(r)}{r}
\label{findr<2a-gel}
\end{eqnarray}
is the microion-induced electrostatic force.

Finally, the density derivative of the effective electrostatic pair potential
between overlapping macroions, which appears in Eq.~(\ref{dUdV}), is given by
\begin{equation}
n_m\left(\frac{\partial v_{\rm eff}(r)}{\partial n_m}\right)_{N_s/N_m}
=\frac{\kappa}{2}\left(\frac{\partial v_{\rm eff}(r)}{\partial\kappa}\right)
\label{dvdn1}
\end{equation}
with
\begin{eqnarray}
&&\beta\left(\frac{\partial v_{\rm eff}(r)}{\partial\kappa}\right)
=-\left(\frac{3Z}{\kt^2}\right)^2\frac{\lambda_B}{2r}\left[
-2+\rt^2+\frac{3}{\kt^2}+\frac{2}{3}\kt^2\rt
\right.
\nonumber\\[1ex]
&&-\frac{1}{2}\kt^2\rt^2+\frac{1}{24}\kt^2\rt^4
+\left(2+\frac{\kappa r}{2}-\frac{3}{\kt^2}-\frac{\rt}{2\kt}\right)e^{-\kappa r}
\nonumber\\[1ex]
&&-\left(4+\kt+\frac{6}{\kt}+\frac{3}{\kt^2}\right) e^{-2\kt}\sinh(\kappa r)
\nonumber\\[1ex]
&&\left.+\frac{1}{2}\left(1+\frac{1}{\kt}\right)^2 e^{-2\kt}\kappa r\cosh(\kappa r)\right]~.
\label{dvdn2}
\end{eqnarray}


\vspace*{1cm}


%

\end{document}